\documentclass[preprint,authoryear,12pt]{elsarticle2}

\usepackage{epsfig}

\usepackage{amssymb}
\usepackage{upgreek}
\usepackage{caption}
\usepackage{threeparttable}
\usepackage{verbatim}
\usepackage{gensymb}
\usepackage{natbib}
\usepackage{graphicx, amsmath}
\usepackage{amsthm,amssymb}
\usepackage{multirow}
\usepackage{longtable}
\usepackage{cite}
\usepackage{amsfonts}
\usepackage{hhline}
\usepackage{subfig}

\usepackage[ps2pdf,%
a4paper=true,%
breaklinks=true,%
colorlinks=true,%
pdfauthor={First Author et al.},%
pdftitle={Template for manuscripts in Advances in Space Research}%
]{hyperref}

\journal{Advances in Space Research}

\begin{document}

\begin{frontmatter}

\title{Photometric Transformation from $RGB$ Bayer Filter System to Johnson-Cousins $BVR$ Filter System}

\author[khu]{Woojin Park}\ead{woojinpark@khu.ac.kr}
\author[khu]{Soojong Pak\corref{cor}}\ead{soojong@khu.ac.kr}
\author[knu]{Hyunjin Shim}
\author[khu]{Huynh Anh N. Le}
\author[snu]{Myungshin Im}
\author[kaist]{Seunghyuk Chang}
\author[ka]{Joonkyu Yu}

\address[khu]{School of Space Research and Institute of Natural Sciences, Kyung Hee University\\
1732 Deogyeong-daero, Giheung-gu, Yongin-si, Gyeonggi-do, 446-701, Korea}
\address[knu]{Department of Earth Science Education, Kyungpook National University,\\
Sangyeok 3-dong, Buk-gu, Daegu, Korea}
\address[snu]{CEOU, Astronomy Program, Department of Physics \& Astronomy, Seoul National University,\\
1, Gwanak-ro, Gwanak-gu, Seoul, Korea}
\address[kaist]{Center for Integrated Smart Sensors, Korea Advanced Institute of Science and Technology(KAIST)\\
Nonhyeon-ro 28-gil 25, Gangnam-gu, Seoul 135-854, Korea}
\address[ka]{Hwasangdae Observatory,\\
615-4, Ahopsari-ro, Naechon-myeon, Hongcheon-gun, Gangwon-do 250-862, Korea}

\cortext[cor]{Corresponding author}

\begin{abstract}
The $RGB$ Bayer filter system consists of a mosaic of $R$, $G$, and $B$ filters on the grid of the photo sensors which typical commercial DSLR (Digital Single Lens Reflex) cameras and CCD cameras are equipped with. Lot of unique astronomical data obtained using an $RGB$ Bayer filter system are available, including transient objects, e.g. supernovae, variable stars, and solar system bodies. The utilization of such data in scientific research requires that reliable photometric transformation methods are available between the systems. In this work, we develop a series of equations to convert the observed magnitudes in the $RGB$ Bayer filter system ($R_B$, $G_B$, and $B_B$) into the Johnson-Cousins $BVR$ filter system ($B_J$, $V_J$, and $R_C$). The new transformation equations derive the calculated magnitudes in the Johnson-Cousins filters ($B_{Jcal}$, $V_{Jcal}$, and $R_{Ccal}$) as functions of $RGB$ magnitudes and colors. The mean differences between the transformed magnitudes and original magnitudes, i.e. the residuals, are $\Delta(B_J-B_{Jcal})$ = 0.064 mag, $\Delta(V_J-V_{Jcal})$ = 0.041 mag, and $\Delta(R_C-R_{Ccal})$ = 0.039 mag. The calculated Johnson-Cousins magnitudes from the transformation equations show a good linear correlation with the observed Johnson-Cousins magnitudes.

\end{abstract}


\end{frontmatter}

\parindent=0.5 cm


\section{Introduction}
\label{sec:intro}

Photometric systems play an important role in the quantitative photometric and spectroscopic study of stars and stellar systems. The Johnson-Cousins $UBVRI$ system is the most popularly used broad-band photometric system, and most of the existing optical data have been observed with this system (see \citealp{bessell2005}).

Lately, the chance of amateur astronomers taking parts in scientific research is increasing especially in fields of transient objects and variable stars. A large amount of the photometric data for either supernovae or variable stars are available with amateur CCD cameras, in which the $RGB$ Bayer filter system is commonly adopted. Since most of the previously compiled photometric data in the astronomical community have been taken with the Johnson-Cousins $UBVRI$ photometric system (\citealp{rodgers2006}; \citealp{stencel2013}), reliable transformation equations are necessary in order to utilize the data from the $RGB$ filter system.

There have been many works on the transformation of $RGB$ photometry to Johnson-Cousins photometry in the literature (\citealp{hoot2007}; \citealp{loughney2010}; \citealp{kloppenborg2012}; \citealp{vitek2012}). Especially, it was shown that the filter response function of Bayer $Green$ filter is comparable to that of Johnson $V$ filter (see \citealp{loughney2010}; \citealp{kloppenborg2012}). \citet{vitek2012} transformed Bayer filter system to Johnson-Cousins filter system using color indices of stars. However, there is no equation to convert $RGB$ photometry to Johnson-Cousins $UBVRI$ photometry which reflects spectral types, luminosity classes and metallicities of different astronomical sources.

A well-known example of the transformation between the different filter systems is the photometric transformation from the Sloan Digital Sky Survey ($SDSS$) filter system (\citealp{fukugita1996}) to the Johnson-Cousins filter system. $SDSS$ provides homogeneous $ugriz$ photometry for objects in a large fraction of the northern sky (\citealp{karaali2005}; \citealp{rodgers2006}; \citealp{chonis2008}; \citealp{bilir2008,bilir2011,bilir2012}; \citealp{yaz2010}). \citet{fukugita1996} and \citet{smith2002} derived an $g$ magnitude estimation for Sloan photometry of the standard stars from \citet{landolt1992}. \citet{bilir2005} obtained a more accurate $g$ magnitude transformation which covers both $(g-r)$ and $(r-i)$ for late-type dwarf stars. They mainly used color-color correlations to derive the equations. Transformation equations are generally dependent on metallicities and luminosity classes (\citealp{karaali2005}; \citealp{rodgers2006}; \citealp{bilir2008,bilir2011,bilir2012}; \citealp{yaz2010}; \citealp{karaali2013}). Recently, \citet{ak2014} derived transformation equations covering a large range of metallicities, e.g. -4 $\le$ [$Fe/H$] $\le$ 0.5 dex for giant stars.
Photometric data with the DSLR $RGB$ filter system may also be transformed, using the same principle as in the $SDSS$ to Johnson-Cousins magnitude conversion.

In this paper, we present our observations of the open clusters IC4665 and M52, and derive the transformation equations between the $RGB$ filter system and the Johnson-Cousins $BVR$ filter system. Circumstances of the observations and the preprocessing methods are described in $\S$~\ref{sec:observation}. $\S$~\ref{sec:results} describes the process of Point Spread Function (PSF) photometry to derive final transformation equations for M52. $\S$~\ref{sec:discussion} shows results of the transformation including the residuals, and confirms the correlations between the observed data and the transformed data using color-magnitude diagrams. We then apply the derived transformation to IC4665 to check the validity of our equations.
\section{Observations and data reduction}
\label{sec:observation}

\subsection{Observations of M52}
M52 is a 60 $\pm$ 10 Myr old open star cluster that has a spectral type of B2 to G2. It has interstellar reddening, E($B$ $-$ $V$) = 0.65 mag (\citealp{bonatto2006}), and has similar solar metallicity which is $[Fe$/$H]$ $=$ 0 (\citealp{tiffany2008}).
$BVR$ data of M52 were taken in 2014 June 27 using a 4k $\times$ 4k CCD camera with 15 $\mu$m pixels with the D = 1000 mm f/7.5 RC-telescope at the Lemmonsan Optical Astronomy Observatory ($LOAO$), USA. The weather was clear, and the atmosphere is stable with a flux variation is only under 2\%, and the seeing disk size of $\sim$3 pixels ($\sim$2.5$^{\prime\prime}$). We obtained 18 images in each band, with 15 sec exposures in $B$, 9 sec exposures in $V$, and 8.5 sec exposures in $R$. The field of view for the $BVR$ data of M52 is 28$^\prime$ $\times$ 28$^\prime$. These data were calibrated using WEBDA\footnote{http://www.univie.ac.at/webda/navigation.html} database (\citealp{pesch1960}; \citealp{haug1970}; \citealp{wramdemark1976}; \citealp{viskum1997}; \citealp{choi1999}; \citealp{stetson2000}; \citealp{pandey2001}; \citealp{maciejewski2007}). The CCD was cooled at  $-110$$^{\circ}$C.

\begin{figure}
\centering
\includegraphics[width=0.8\textwidth]{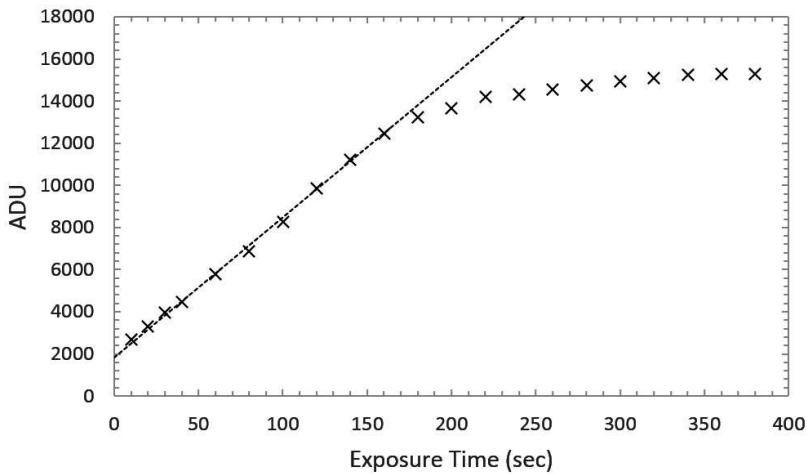}
\caption{\label{550Dlinearity} \scriptsize A linearity plot for the EOS 550D camera. The dashed line represents linearity and it fits the curve well up to 13000 ADU. The full-well capacity is 15300 ADU. The data were obtained by taking flat images.}
\end{figure}

$RGB$ data of M52 were taken in 2014 October 09 with an IR-cut-filter-removed EOS 550D (Canon Inc.\footnote{http://www.canon.com}) camera on the D = 300 mm f/8 RC-telescope at Hwasangdae Observatory ($HSDO$), Hongcheon-gun, Gangwon-do, Korea. The telescope is equipped with a field corrector, which extends the focal length slightly. The effective focal length of the telescope is 2600 mm.
The EOS 550D camera we have used has an APS-C CMOS sensor (sized 22.3 $\times$ 14.9 mm) with 4.3 $\mu$m pixels. We selected a non-process mode (RAW images), and a low amplification mode (ISO = 400). The full-well capacity of the EOS 550D's CMOS sensor is about 15300 ADU, while the linearity is guaranteed up to 13000 ADU at ISO = 400 (Figure~\ref{550Dlinearity}). Since the background noise increases dramatically as battery charge decreases or after changing the batteries (\citealp{henden2014}), we used an AC adapter instead of batteries.
The field of view for the $RGB$ data of M52 is 23$^\prime$ $\times$ 30$^\prime$. The data was obtained by taking 21 images, 120 sec exposures each at 4.7 pixels ($\sim$1.6$^{\prime\prime}$) seeing. Note that the Full Width at Half Maximum (FWHM) of the DSLR images should be larger than 2 pixels to cover all three (Red, Green, and Blue) pixels with the Bayer filter set.

\subsection{Observations of IC4665}
IC4665 is a 27.7 $\pm$ 1.1 Myr old open star cluster with metallicity of $[Fe$/$H]$ $=$ $-$0.03 $\pm$ 0.04 (\citealp{manzi2008}; \citealp{jeffries2009}; \citealp{smith2011}; \citealp{lodieu2011}). It has interstellar reddening, E($B$ $-$ $V$) = 0.18 mag (\citealp{prosser1993}).
$BVR$ magnitudes of stars in IC4665 are obtained from the WEBDA database (\citealp{johnson1954}; \citealp{hogg1955}; \citealp{alcaino1965}; \citealp{mccarthy1969}; \citealp{eggen1971}; \citealp{sanders1972}; \citealp{muzzio1973}; \citealp{neckel1974}; \citealp{landolt1983a}; \citealp{landolt1983b}; \citealp{menzies1991}; \citealp{prosser1993}; \citealp{menzies1996}; \citealp{de wit2006}).

$RGB$ data of IC4665 were obtained in 2014 April 14. The observations were made with the EOS 550D camera attached to a William Optics ZenithStar 80 ¥± ED APO telescope (D = 80mm, f/6.8, William Optics Corporation\footnote{http://www.williamoptics.com}) at the Kyung Hee Astronomical Observatory ($KHAO$), Gyeonggi-do, Korea. The camera settings were identical to those of the M52 observations.
The field of view ($FOV$) of the system is 94$^\prime$ $\times$ 141$^\prime$. We took 30 frames with 15 sec exposures during a photometric night with the seeing disk size of $\sim$5 pixels ($\sim$8.1$^{\prime\prime}$).

The details of the observations, including filters, target coordinates, field of views, and observing conditions are summarized in Table~\ref{table:log}.

\scalebox{0.65}{
\setlength{\LTleft}{-1cm}
\setlength{\LTright}{\LTleft}
    \begin{threeparttable}
    \begin{longtable}{cccccccc}
    \captionsetup{justification=raggedleft}
    \caption{Observation log} \\
    \hhline{========}
    Date & Observatory and Telescope & Filters\tnote{e} & Target & R.A. & Dec.(J2000) & FoV\tnote{a} & Seeing ($^{\prime\prime}$)\\
    \hline
    \endhead
    2014 Apr 14	&	KHAO\tnote{b}, 0.08m telescope	& $RGB$  &IC4665	  &	 17h46m18s	    &	 $+05^\circ 43^\prime 00^{\prime\prime}$   &	94$^\prime$ $\times$ 141$^\prime$ & 8.1	\\
    2014 Jun 27	&	LOAO\tnote{c}, 1m    telescope 	& $BVR$  &M52         &	 23h24m48s	    &	 $+61^\circ 35^\prime 36^{\prime\prime}$   &	28$^\prime$ $\times$ 28$^\prime$  &	 2.5 \\
    2014 Oct 09	&	HSDO\tnote{d}, 0.3m  telescope	& $RGB$  &M52         &	 23h24m48s	    &	 $+61^\circ 35^\prime 36^{\prime\prime}$	&	23$^\prime$ $\times$ 30$^\prime$  & 1.6 \\
    \hhline{========}
    \label{table:log}
    \end{longtable}
    \begin{tablenotes}
    \item[a]{\footnotesize Field of View}
    \item[b]{\footnotesize Kyung Hee Astronomical Observatory, http://khao.khu.ac.kr}
    \item[c]{\footnotesize Lemmonsan Optical Astronomy Observatory, http://loao.kasi.re.kr}
    \item[d]{\footnotesize Hwasangdae Observatory, 915-2 Hwasangdae-ri, Naechon-myeon, Hongcheon-gun, Gangwon-do, Korea}
    \item[e]{\footnotesize $RGB$ is the Bayer $RGB$ filter system for the Canon EOS 550D, and $BVR$ is the Johnson-Cousins filter system}
    \end{tablenotes}
    \end{threeparttable}
    }

\subsection{Data reduction}

\begin{figure}
\centering
\includegraphics[width=0.8\textwidth]{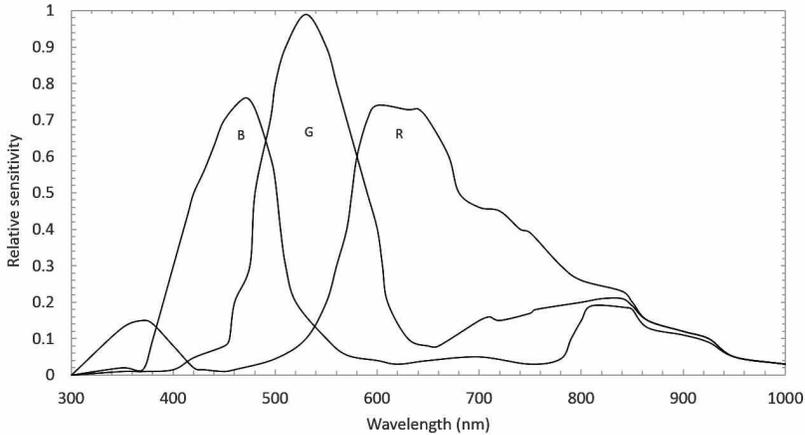}
\caption{\label{rgbcurve} \scriptsize Throughput of Bayer $R$, $G$, and $B$ filter for the EOS 550D (\citealp{cerny2013}). The vertical axis is relative sensitivity for these filters.}
\end{figure}

The $RGB$ Bayer filter system consists of 2 $\times$ 2 pixels mosaic $RGB$ filter set that covers the sensor with $Red$, $Green$, $Green$, and $Blue$ colors (\citealp{tweet2009}), thus we need to separate the single image into images with different filters. We wrote a python-based program to extract pixels that correspond to each color to produce $R$, $G$, and $B$ filter images. Especially for $Green$ filters, we combined two $Green$ filter images by averaging for each set of $RGGB$ filters. $R$, $G$, and $B$ filters have center wavelength, $\lambda_{eff}$ = 640 nm, 530 nm, and 470 nm, respectively. Figure~\ref{rgbcurve} shows throughput of $R$, $G$, and $B$ filters for the EOS 550D (\citealp{cerny2013}). Since the Bayer filters are directly attached on the CMOS substrate, the throughput shows the convolution of the filter transmissions and the quantum efficiency of the detector. Image combining and basic preprocessing were performed with the Image Reduction and Analysis Facility ($IRAF$)\footnote{IRAF is distributed by the National Optical Astronomy Observatory, which is operated by the Association of Universities for Research in Astronomy, Inc., under cooperative agreement with the National Science Foundation.}.

Even though the CMOS of the DSLR camera was not cooled, the average dark current for sufficiently short exposures ($\sim$120 seconds) was low, i.e. 1 - 3 ADU. As a result of our dark current test, we expect that the Canon image processor, DIGIC 4, automatically correct dark currents. Therefore we did not apply dark subtraction to the object images. We performed bias and flat corrections before we break images into sub-frames, then we combined 30 images for each filter to make a single object image to reduce the random noise.

\section{Results}
\label{sec:results}

\subsection{Photometric data}

After creating the combined images, we derived magnitudes for each star in the star cluster in each filter with the IRAF/DAOPHOT package. Either moffat15 function or penny2 function was chosen as the PSF. We detected 564 stars at 4$\sigma$ over sky background in the M52 $RGB$ image. We removed field stars that do not belong to the cluster using color-magnitude and color-color correlations to get only star cluster members (\citealp{pesch1960}). After identifying the same stars in $BVR$ images, we compared mag($Blue$) with mag($B$), mag($Green$) with mag($V$), and mag($Red$) with mag($R$) for individual stars to derive the conversion formula that describes the correlation between the two filter systems. Based on the color-magnitude diagram, we separate giants, so that only main-sequence stars are selected for transformation.
All magnitudes are de-reddened applying interstellar reddening, $A_V$ = 3.1$\times$$E(B - V)$. Extinction magnitudes for the $RGB$ Bayer filter system are derived by the extinction law (\citealp{cardelli1989}; \citealp{odonnell1994}; \citealp{schlegel1998}). The radius of the aperture for photometry was 9.4 pixels ($\sim$3.2$^{\prime\prime}$) for the $RGB$, and 6 pixels ($\sim$5$^{\prime\prime}$) for the $BVR$ data. Stars in M52 have magnitudes and colors in the range of 8.49 $\le$ $V$ $\le$ 14.14 mag and -0.15 $\le$ $B-V$ $\le$ 0.48 mag, respectively.

\subsection{Transformation equations}
\label{subsec:transformation}
The transformation equations are described as follows:
\begin{align}
    B_J &= B_{B,ZP}+B_B+C_{B,BG}(B_B-G_B)+C_{B,GR}(G_B-R_B) \\
    V_J &= G_{B,ZP}+G_B+C_{V,BG}(B_B-G_B)+C_{V,GR}(G_B-R_B) \\
    R_C &= R_{B,ZP}+R_B+C_{R,BG}(B_B-G_B)+C_{R,GR}(G_B-R_B)
\label{eq1}
\end{align}
In these equations, $B_J$, $V_J$, and $R_C$ are the Johnson $B$, $V$, and Cousins $R$ magnitudes, respectively. Also, $B_B$, $G_B$, and $R_B$ indicate Bayer $B$, $G$, and $R$ magnitudes. In addition, we include $B_{B,ZP}$, $G_{B,ZP}$, and $R_{B,ZP}$ representing the zero-points for the Bayer filters. Zero-points vary depending on the observing conditions such as weather, air mass, and the system sensitivity.
The coefficients were derived by the iterative least square fitting algorithm with sigma clipping method (see Table~\ref{table:coeffi}).

\scalebox{0.9}{
\setlength{\LTleft}{2cm}
\begin{threeparttable}[h]
\begin{centering}
\begin{longtable}[h]{lccc}
\captionsetup{justification=centerlast}
\caption{Coefficients and $RMS$ variation by Sigma Clipping method} \\
\hline
&Original Data & $\mathrm{1^{st}}$ SC\tnote{a} & $\mathrm{2^{nd}}$ SC\tnote{a} \\
\hline
\endhead
$B_{B,ZP}$  &	-0.371 &	-0.286	&	$-0.291\substack{+0.026 \\ -0.028}$	 \\
$C_{B,BG}$	&	0.246	&	0.419	&	$0.280\substack{+0.318 \\ -0.292}$	 \\
$C_{B,GR}$	&	0.213	&	0.543	&	$0.600\substack{+0.133 \\ -0.118}$	 \\
\hline
$RMS$       &	0.220	&	0.083	&	0.064	\\
\hline\hline
$G_{B,ZP}$	&	-0.322	&	-0.263	&	$-0.252\substack{+0.022 \\ -0.025}$	 \\
$C_{V,BG}$	&	0.578	&	0.713	&	$0.542\substack{+0.273 \\ -0.263}$	 \\
$C_{V,GR}$	&	-0.471	&	-0.196	&	$-0.064\substack{+0.119 \\ -0.103}$	 \\
\hline
$RMS$       &	0.185	&	0.057   	&	0.041	\\
\hline\hline
$R_{B,ZP}$	&	-0.277	&	-0.240	&	$-0.226\substack{+0.031 \\ -0.031}$	 \\
$C_{R,BG}$	&	0.123	&	0.226	&	$0.051\substack{+0.334 \\ -0.368}$	 \\
$C_{R,GR}$	&	0.068	&	0.298	&	$0.468\substack{+0.142 \\ -0.145}$	 \\
\hline
$RMS$       &	0.163	&	0.055	&	0.039	\\
\hline\hline
\label{table:coeffi}
\end{longtable}
\begin{tablenotes}
\item[a]{\footnotesize Sigma Clipping method with 2.5$\sigma$}
\end{tablenotes}
\end{centering}
\end{threeparttable}
}
\vspace{1\baselineskip}

\begin{figure}
\centering
\includegraphics[width=0.8\textwidth]{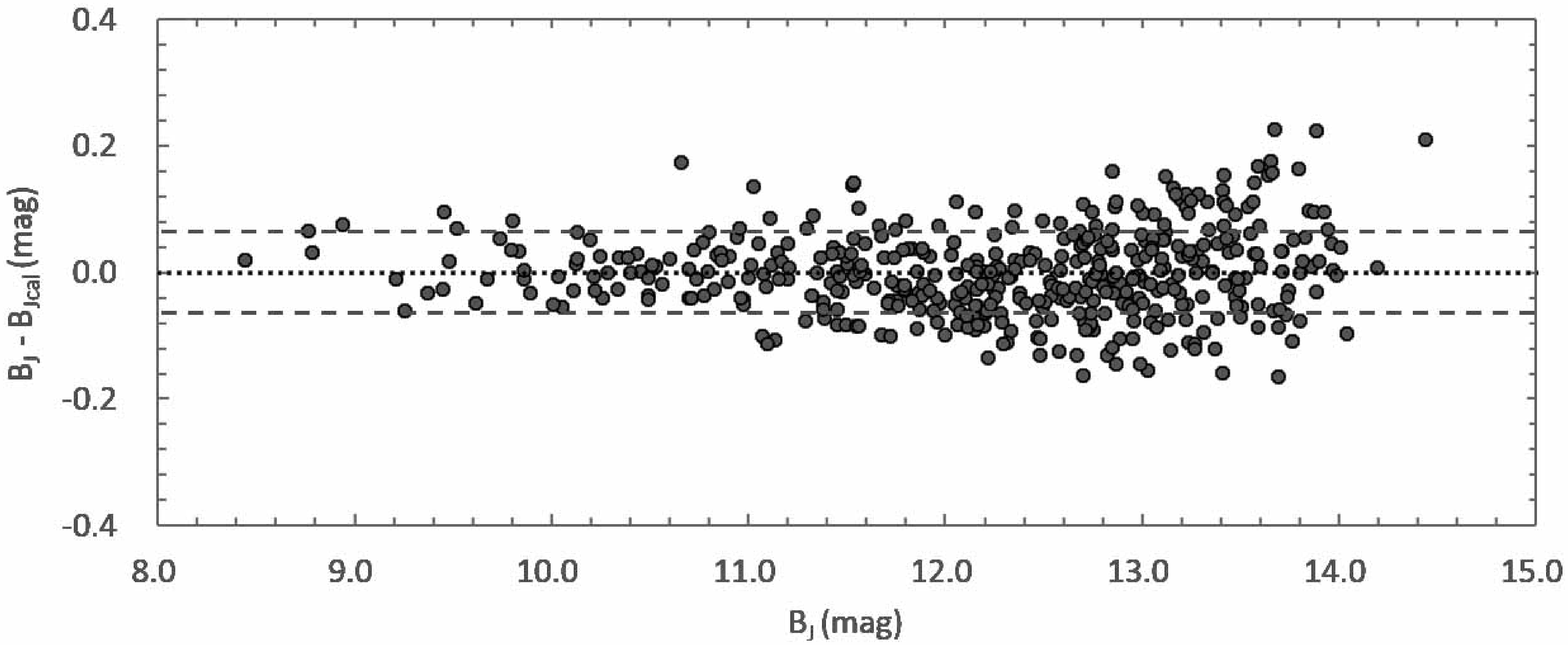}
\includegraphics[width=0.8\textwidth]{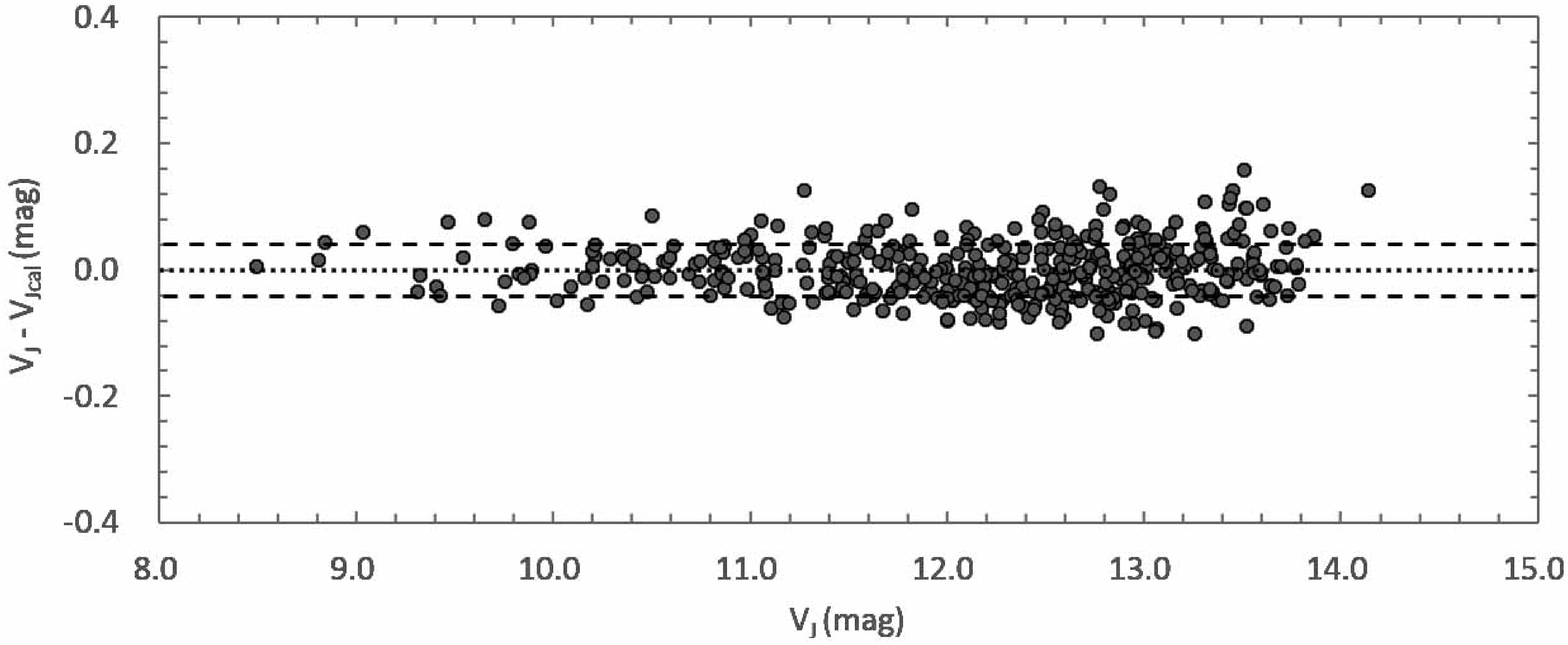}
\includegraphics[width=0.8\textwidth]{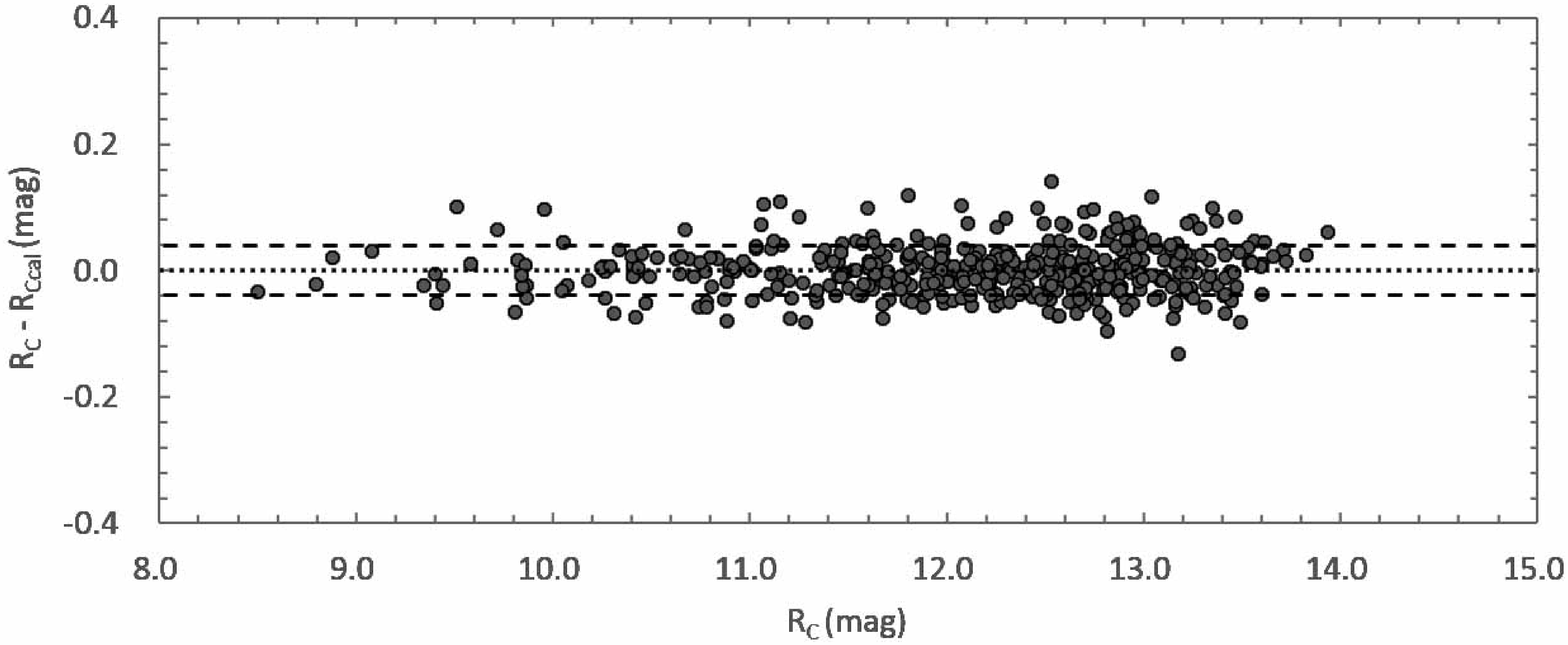}
\caption{\label{M52residuals}\scriptsize Residuals plots for the $B_J$, $V_J$, and $R_C$ bands for M52 from top to bottom. The horizontal axis indicates the observed Johnson-Cousins magnitudes. Dashed lines represent the position of 1$\sigma$ dispersions, $\pm$0.064, $\pm$0.041, and $\pm$0.039 for the $B_J$, $V_J$, and $R_C$ bands.}
\end{figure}

Figure~\ref{M52residuals} shows the distribution of magnitude residuals for the Johnson-Cousins systems, i.e. the difference between the observed magnitudes and the calculated magnitudes based on the transformation equations. Most of the objects fall near the zero residual line, which confirms the reliability of our transformation equations. However, there are a number of outliers especially in $B$ band.

\section{Discussion}
\label{sec:discussion}

\subsection{Accuracy of the $RGB$ transformation}
\label{subsec:discussion1}
We derived Root Mean Square ($RMS$) values to test the accuracy of the transformation equations. The $RMS$ errors for M52 are 0.064, 0.041, and 0.039 mag for $B_J$, $V_J$ and $R_C$, respectively (Table~\ref{table:coeffi}).

As Figure~\ref{M52residuals} shows the dispersion around the zero line is the smallest for $R_C$ while it is the largest for $B_J$. The $RMS$ values also suggest that the conversion into $R_C$ magnitudes is more reliable than that of the $B_J$ magnitudes.

\begin{figure}
\centering
\includegraphics[width=0.45\textwidth]{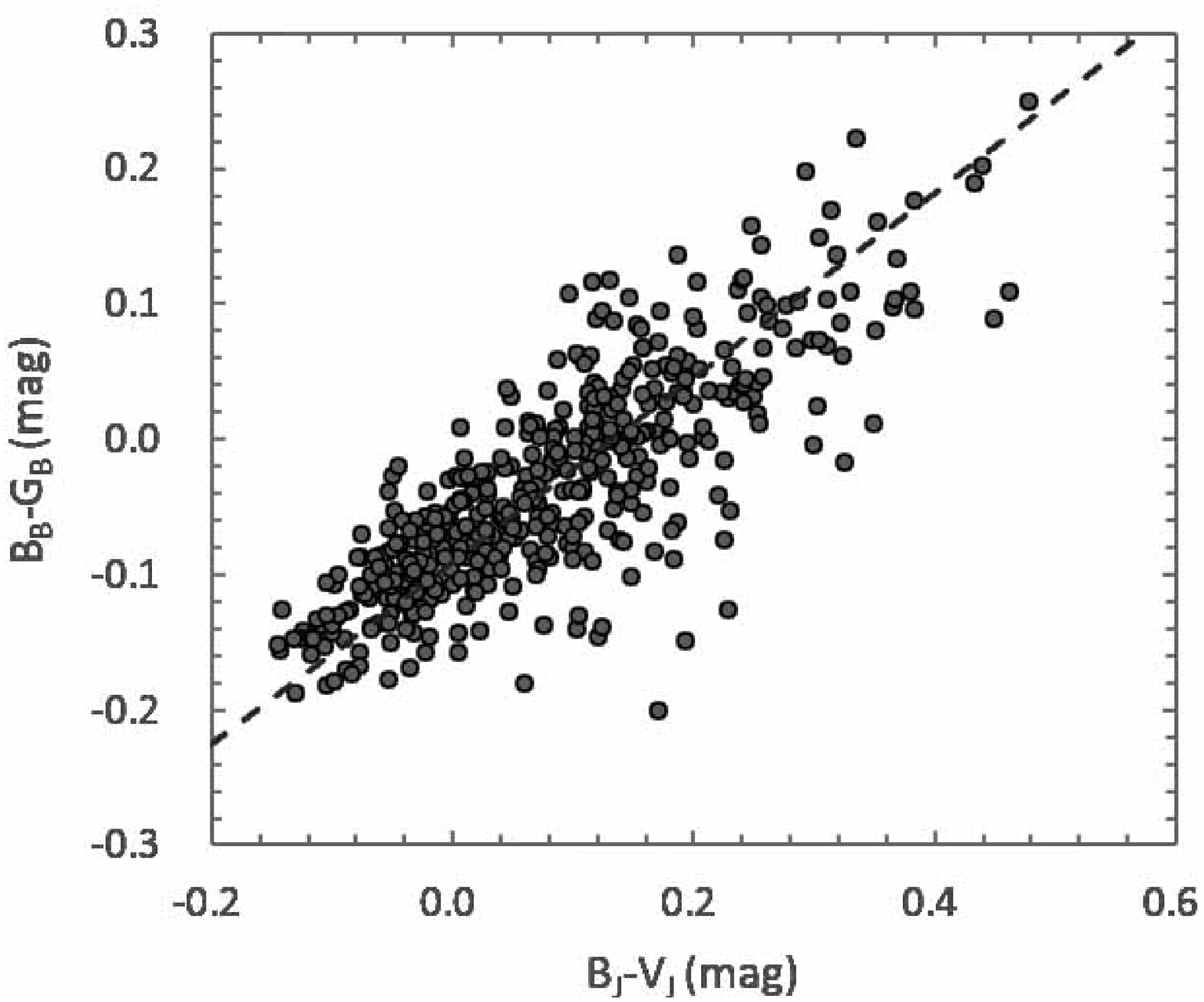}
\includegraphics[width=0.45\textwidth]{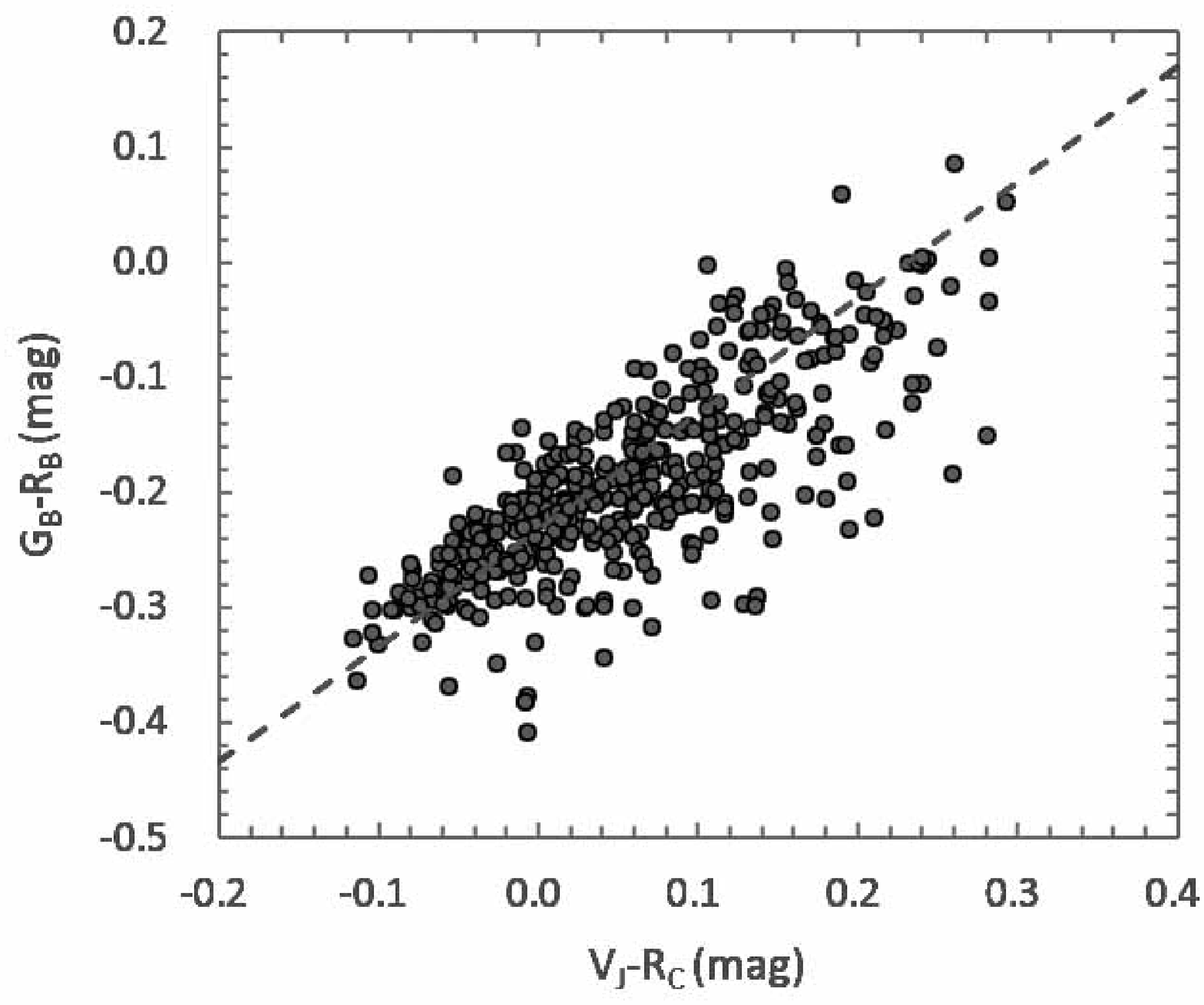}
\caption{\label{M52Colordiagram}\scriptsize Diagram of $RGB$ colors versus $BVR_C$ colors for stars in M52. The dashed lines represent $\mathrm{1^{st}}$ order linear fitting lines.}
\end{figure}

The distributions of the data in two color-color planes corresponding to two systems in Figure~\ref{M52Colordiagram} are linear, and the dispersions are relatively small which indicate that our transformations are reliable. The $RMS$ errors may arise from the contamination of absorption lines and from the interference among wavelength coverage of Bayer $R$, $G$, and $B$ bands.

\begin{figure}[h]
\centering
\includegraphics[width=0.45\textwidth]{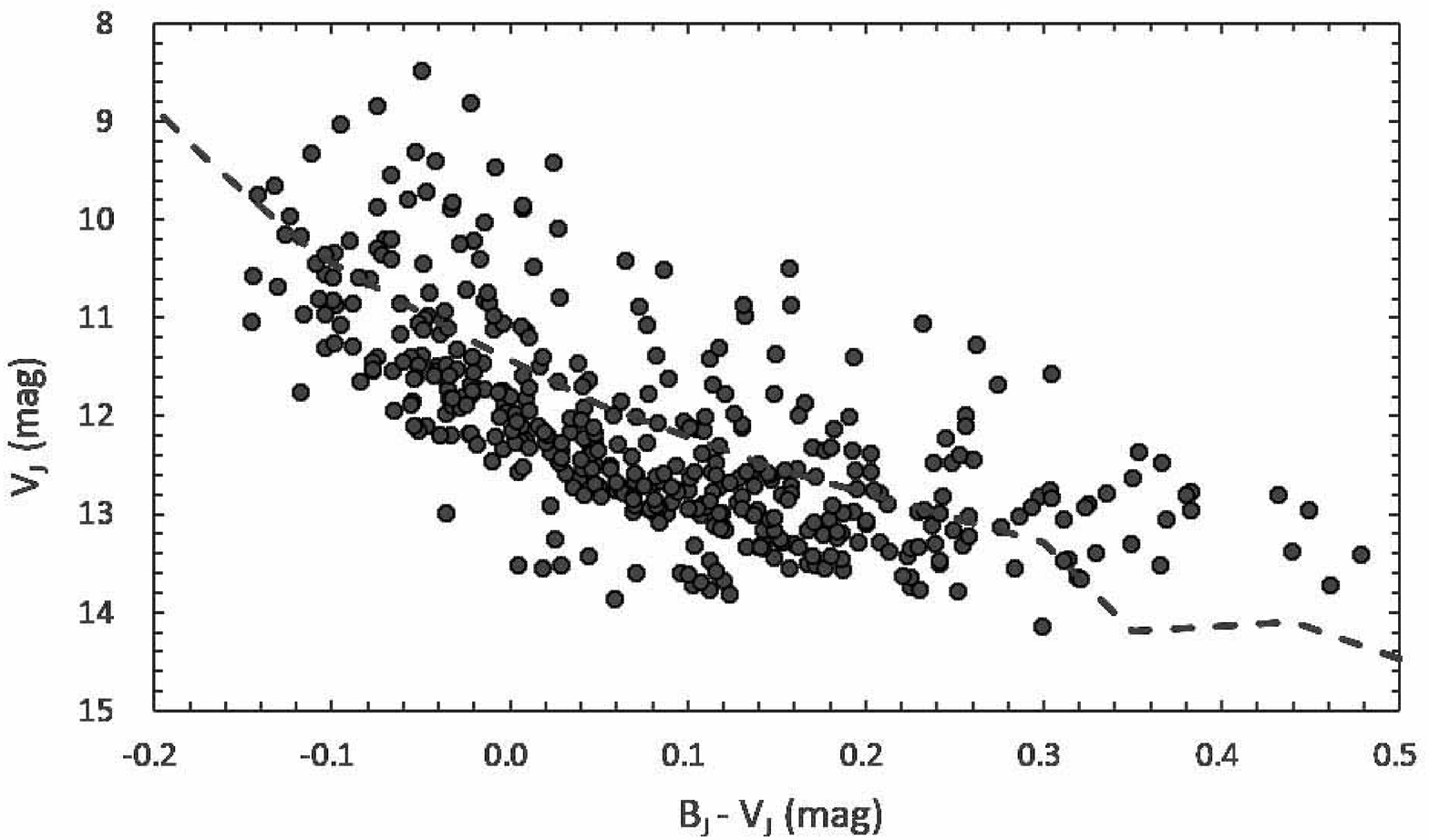}
\includegraphics[width=0.45\textwidth]{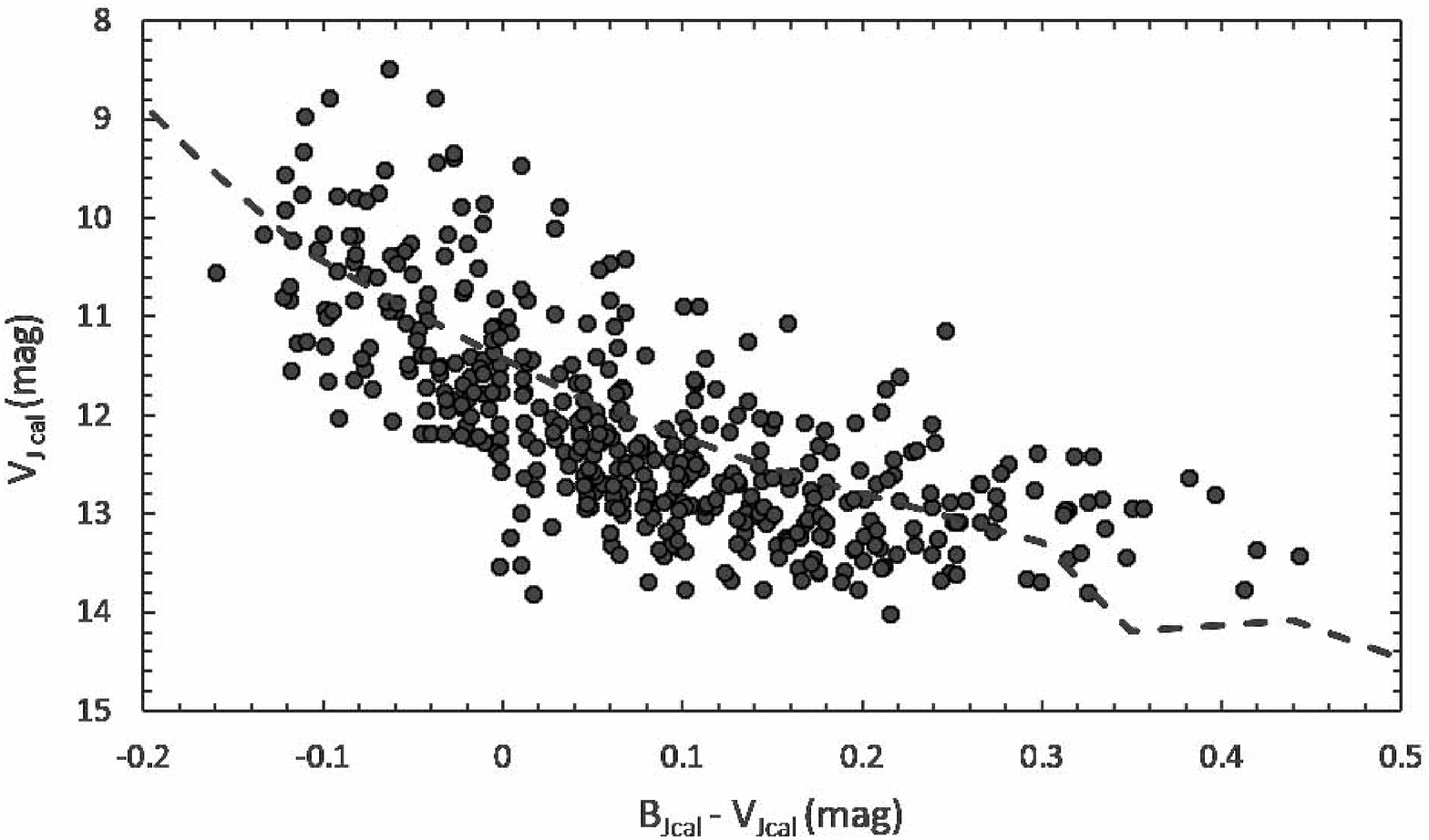}
\caption{\label{M52HRdiagram}\scriptsize M52 color-magnitude diagrams from observed data (top) and calculated data (bottom). The dashed line represents the main-sequence (\citealp{cox2000}). The color-magnitude diagram based on the magnitudes from our transformation (bottom) is consistent with that of the observed magnitudes (top).}
\end{figure}

In Figure~\ref{M52HRdiagram}, we plot the color-magnitude diagrams of M52. All of the stars in M52 fall onto the main-sequence. The similarity of the color magnitude diagrams constructed using the observed Johnson-Cousins filter system magnitudes and the transformed magnitudes shows that the transformation equations work well.

Coefficients in our results have uncertainties derived from the reduced chi-squared method.
\begin{align}
{\chi}^2_{red} = \frac{1}{n-1} \sum\limits_{i}^n \frac{(O_i-E_i)^2}{{\sigma_i}^2},
\label{eq2}
\end{align}
where $n$ is the number of stars, $O$ is the observed data and $E$ is the calculated data, and $\sigma$ is the error the observed data and the calculated data. Derived uncertainties by the Equation (\ref{eq2}) are indicated in Table~\ref{table:coeffi}.

\begin{figure}
\centering
\includegraphics[width=0.8\textwidth]{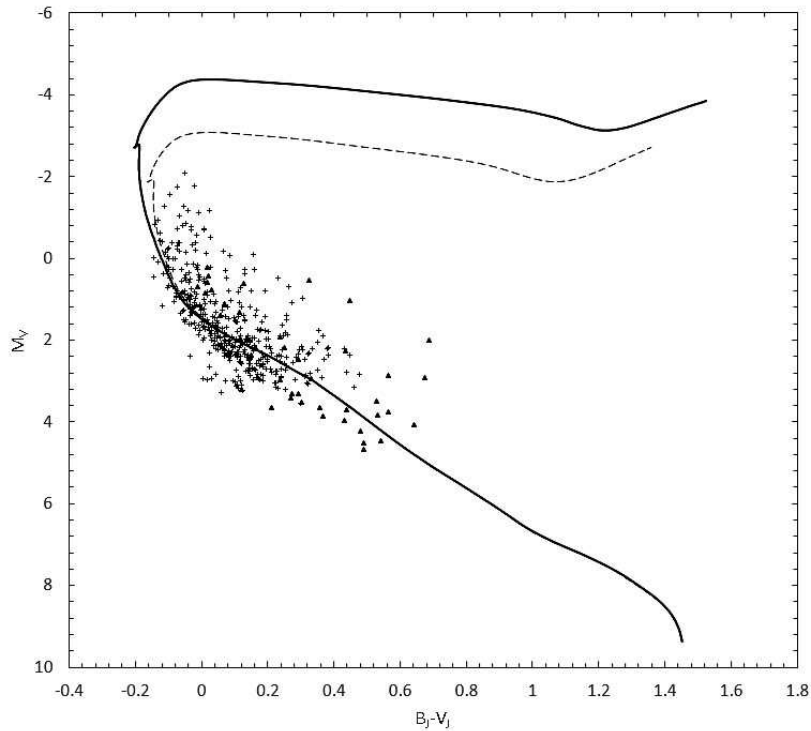}
\caption{\label{isochrone}\scriptsize Isochrones of M52 and IC4665 (\citealp{cassisi2006}). The dashed line and cross symbols indicate the isochrone and observed data points of M52. And the solid line and triangle symbols indicate the isochrone and observed data points of IC4665. M52 and IC4665 have similar metallicities with different ages.}
\end{figure}

\subsection{Application to IC4665}

We applied our transformation to the $RGB$ magnitudes of IC4665. IC4665 is a younger open cluster, which consists of main-sequence stars and sub-giants. We excluded sub-giants for transforming stars as our transformation is focused on main-sequence stars. 

\begin{figure}
\centering
\includegraphics[width=0.8\textwidth]{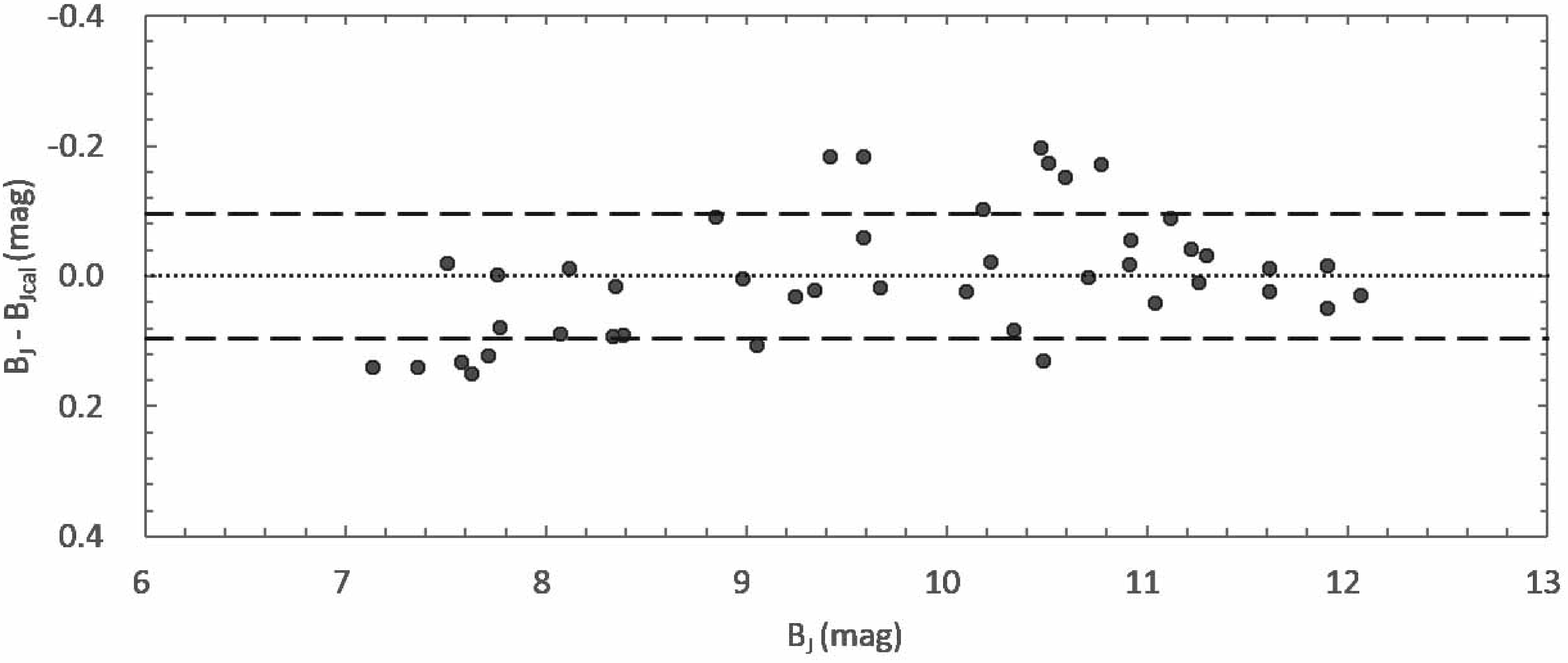}
\includegraphics[width=0.8\textwidth]{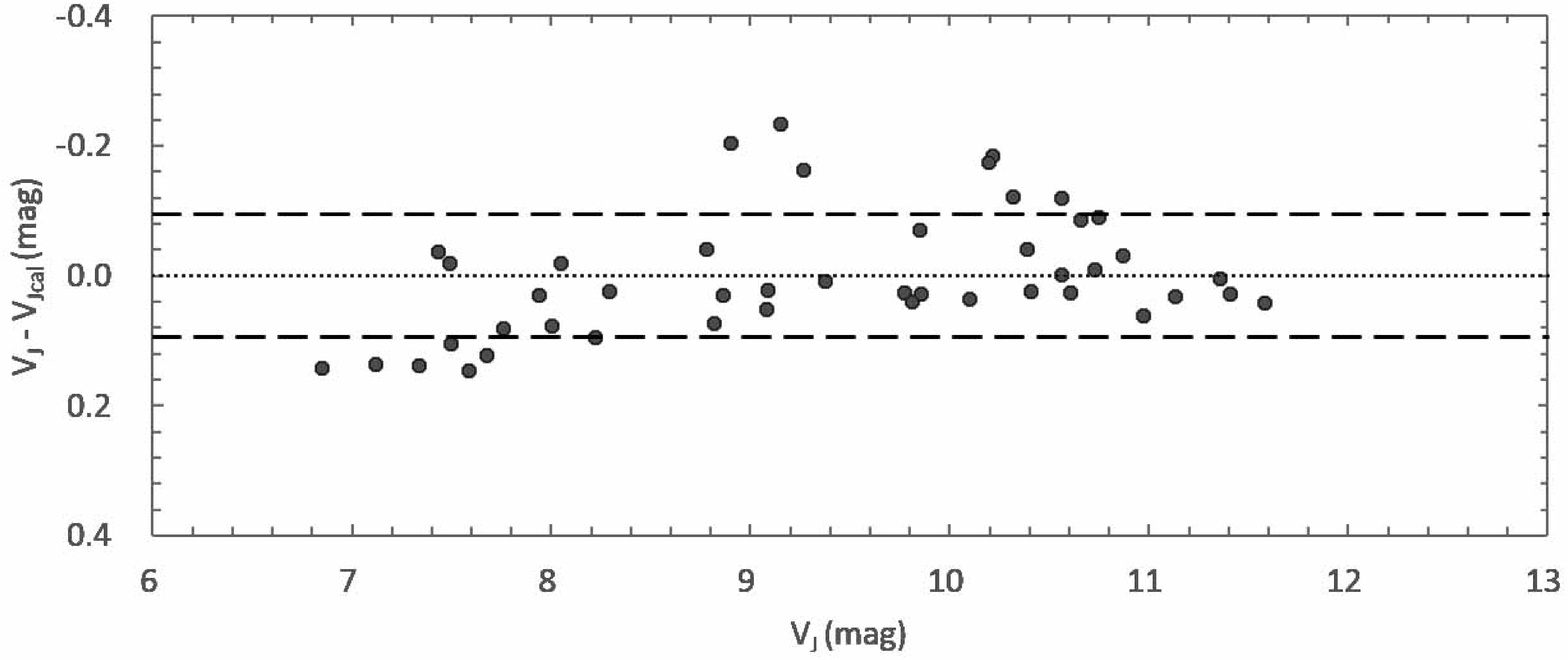}
\includegraphics[width=0.8\textwidth]{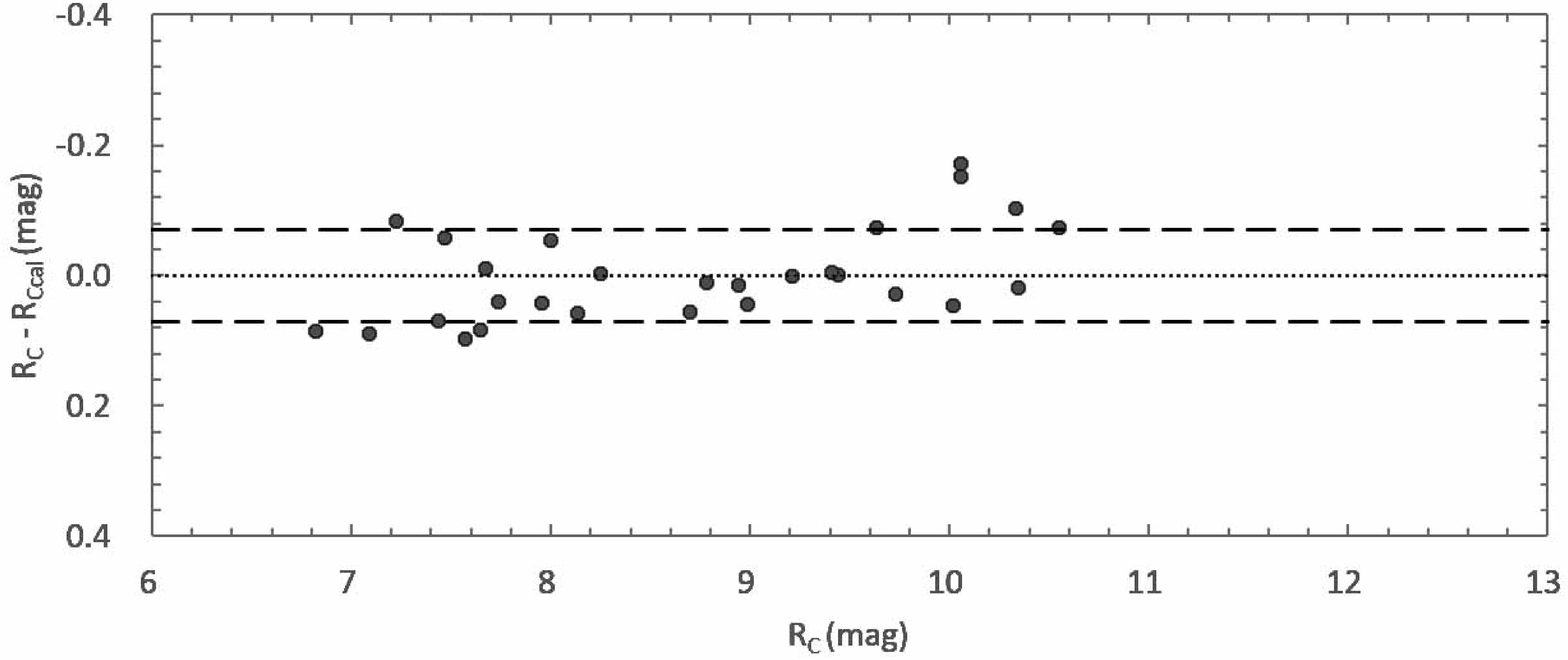}
\caption{\label{IC4665residuals}\scriptsize Residuals plots for the $B_J$, $V_J$, and $R_C$ bands for IC4665 from top to bottom. The horizontal axis is the measured Johnson-Cousins magnitude. Dashed lines represent the position of 1$\sigma$ dispersions, $\pm$0.095, $\pm$0.094, and $\pm$0.070 for the $B_J$, $V_J$, and $R_C$ bands.}
\end{figure}

\begin{figure}
\centering
\includegraphics[width=0.45\textwidth]{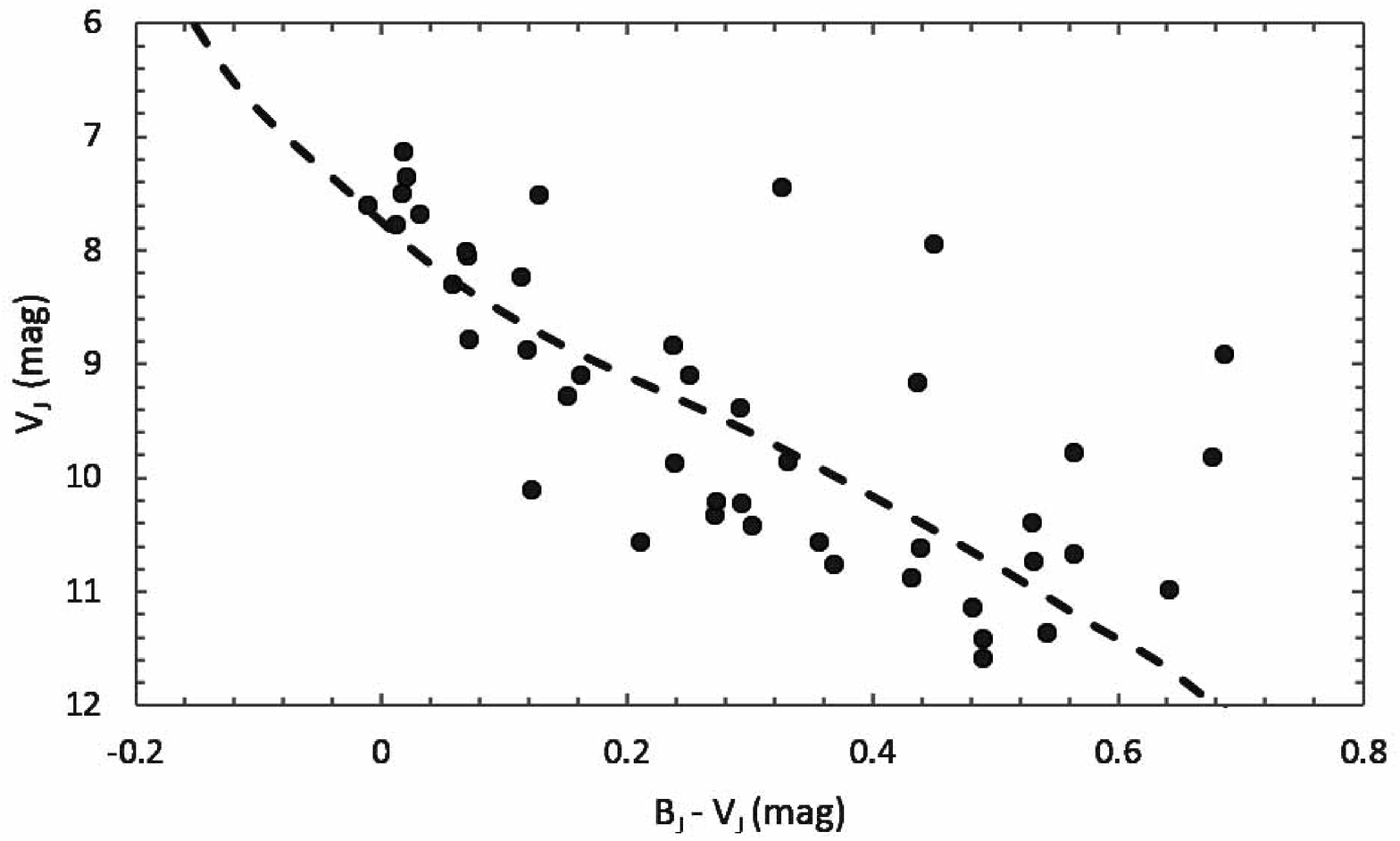}
\includegraphics[width=0.45\textwidth]{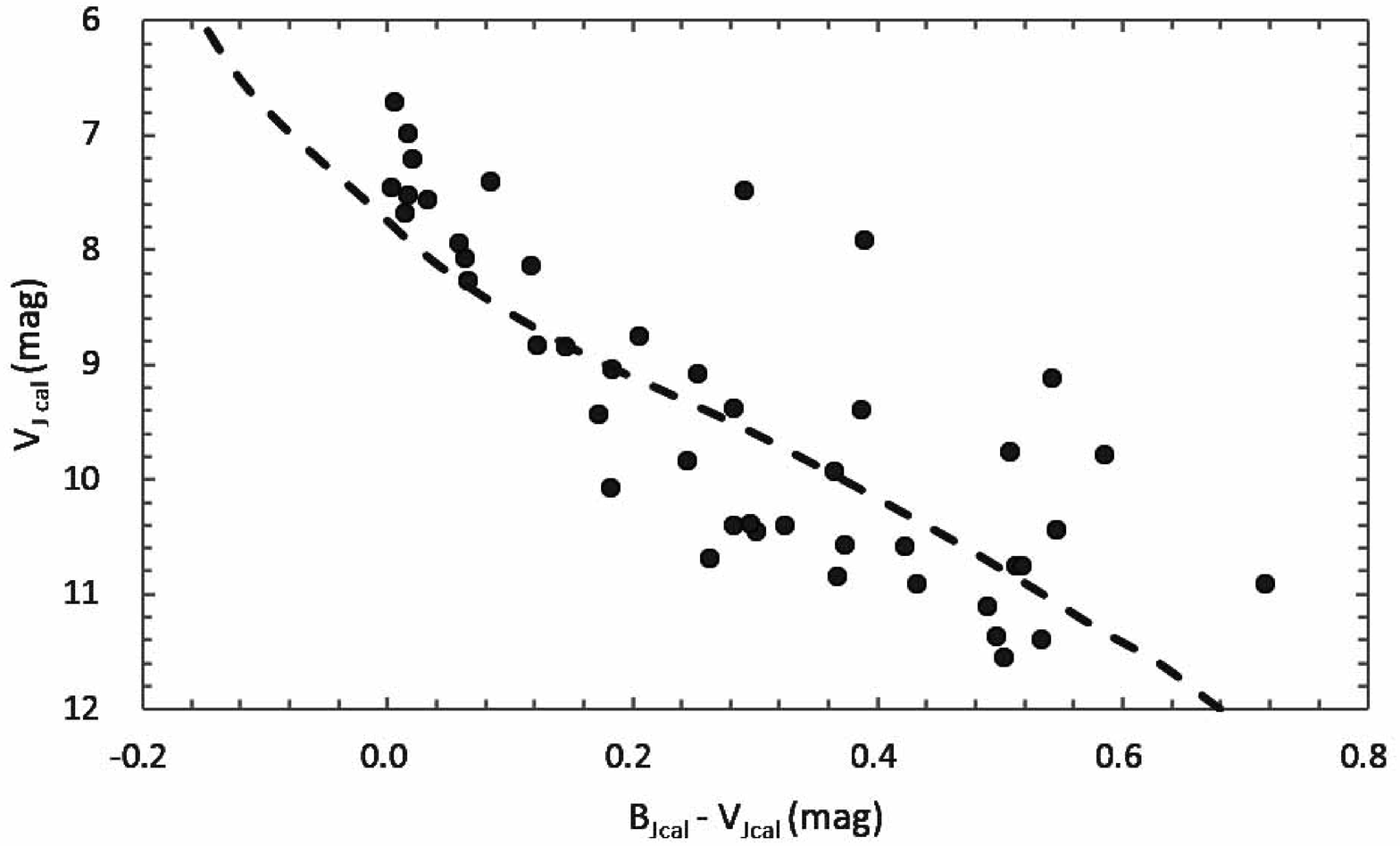}
\caption{\label{IC4665HRdiagram}\scriptsize IC4665 color-magnitude diagrams from the observed data (top) and from the calculated magnitudes (bottom). The dashed line represents the main-sequence. The calculated data were derived using the coefficients estimated from M52. Only the zero-points adjusted for IC4665. The color-magnitude diagram based on calculated data (bottom) are consistent with that of the observed Johnson $B$ and $V$ magnitudes (top).}
\end{figure}

Figure~\ref{isochrone} shows isochrones of M52 and IC4665. Both have similar metallicities with different ages. Since the isochrones of M52 and IC4665 are similar, the filter transformation coefficients can be same. The $RGB$ magnitudes of stars in IC4665 were also measured using IRAF/DAOPHOT with the same method we used for M52.

45 stars were detected with 4$\sigma$ in main-sequence stars, and their $RGB$ magnitudes were transformed into $BVR$ magnitudes using the equations derived in $\S$~\ref{subsec:transformation}. All coefficients in Equations (1) - (3) are kept the same except the zero points, because the observing conditions of IC4665 are different from those of M52.
We select the offset zero points at which the average residual becomes zero. The zero points of IC4665 are -5.813, -5.400, and -5.086 for $B$, $V$, and $R$ bands.

The $RMS$ errors for IC4665 are 0.095, 0.094, and 0.070 mag for $B_J$, $V_J$ and $R_C$, respectively. Figure~\ref{IC4665residuals} shows the residuals of IC4665 between observed magnitudes and transformed magnitudes.

Figure~\ref{IC4665HRdiagram} shows the color magnitude diagrams of IC4665 constructed using the observed magnitudes from WEBDA (top), and the Johnson-Cousins magnitudes calculated from the $RGB$ magnitudes (bottom). The color-magnitude diagrams from the transformed magnitudes follow the same trends as the diagrams from the observed $BVR$ values. The similarity shows that the coefficients work well not only for M52 but also for other clusters.

\section{Summary}
\label{sec:summary}

Equations (1), (2), and (3) convert the magnitudes $RGB$ Bayer filter system to the ones in Johnson-Cousins filter system. The color terms, ($B_B$ $-$ $G_B$) and ($G_B$ $-$ $R_B$) are included in the transformation equations. The coefficients are listed in Table~\ref{table:coeffi}.

There have been a number of previous attempts for filter transformation (e.g. \citealp{rodgers2006}; \citealp{chonis2008}; \citealp{smith2002}; \citealp{karaali2013}; \citealp{ak2014}). The best instrumental fits before our research was obtained by \citet{hoot2007}, who presented $RMS$ errors of 0.347, 0.134 and 0.236 mag for $B$, $V$, and $R$ bands, respectively. Our results are more reliable than those of \citet{hoot2007}, because of the use of two color terms in the transformation equations.

Note that our work is based on a single star cluster, which contains relatively young stars. Although the equations can be applied to other open cluster data (e.g. IC4665), the magnitude conversion would be deviated in other spectral types or metallicities. Further works are needed in order to construct more accurate magnitude transformation equations between stars, galaxies, supernovae, and other objects. \\

\section*{Acknowledgements}
This work was supported by the National Research Foundation of Korea ($NRF$) grant, No. 2008-0060544, funded by the Korean government ($MSIP$). This paper includes data taken at Lemmonsan Optical Astronomy Observatory operated by the Korean Astronomy and Space Science Institute, Kyung Hee Astronomical Observatory operated by Kyung Hee University, and Hwasangdae Observatory. This research used the WEBDA database, operated at the Department of Theoretical Physics and Astrophysics of the Masaryk University. The author would like to thank to Dr. Judit Gyorgyey Ries, UT/McDonald Observatory for editing the English version of the paper.


\begin{thebibliography}{}

\bibitem[Ak et al.(2014)]{ak2014}Ak, S., Ak, T., Karaali, S., et al., 2014. Colour Transformations Between $BVRc$ and $g^\prime r^\prime i^\prime$ Photometric Systems for Giant Stars, PASA 31, 14.
\bibitem[Alcaino(1965)]{alcaino1965}Alcaino, G., 1965. A Photometric Investigation of the Galactic Clusters IC4665 and IC4756, Lowell Obs. 6, 167.
\bibitem[Bessell(2005)]{bessell2005}Bessell, M. S., 2005. Standard Photometric System, ARA\&A 43, 293.
\bibitem[Bilir et al.(2005)]{bilir2005}Bilir, S., Karaali, S., Tun\c{c}el, S., 2005. Absolute Magnitudes for Late-type Dwarf Stars for Sloan Photometry, AN 326, 321.
\bibitem[Bilir et al.(2008)]{bilir2008}Bilir, S., Ak, S., Karaali, S., et al., 2008. Transformations Between 2MASS, SDSS and $BVRI$ Photometric Systems: Bridging the Near-Infrared and Optical, MNRAS 384, 1178.
\bibitem[Bilir et al.(2011)]{bilir2011}Bilir, S., Karaali, S., Ak, S., et al., 2011. Transformations Between the $WISE$, 2MASS, SDSS and $BVRI$ Photometric Systems - I. Transformation Equations for Dwarfs, MNRAS 417, 2230.
\bibitem[Bilir et al.(2012)]{bilir2012}Bilir, S., Karaali, S., Da\u{g}tekin, N. D., et al., 2012. Transformations Between WISE, and 2MASS, SDSS, BVI Photometric Systems: II. Transformation Equations for Red Clump Stars, PASA 29, 121.
\bibitem[Bonatto and Bica(2006)]{bonatto2006}Bonatto, C., Bica, E., 2006. Methods for Improving Open Cluster Fundamental Parameters Applied to M52 and NGC3960, A\&A 455, 931.
\bibitem[Cardelli et al.(1989)]{cardelli1989}Cardelli, J. A., Clayton, G. C., Mathis, J. S., 1989. The Relationship Between Infrared, Optical, and Ultraviolet Extinction, ApJ 345, 245.
\bibitem[Cassisi et al.(2006)]{cassisi2006}Cassisi, S., Pietrinferni, A., Salaris, M., et al., 2006. BASTI: An Interactive Database of Updated Stellar Evolution Models, MSAI 77, 71.

\bibitem[Cerny(2013)]{cerny2013}Cerny, J., 2013. System for Captureing Scene and NIR Relighting Effects in Movie Postproduction Transmission, PCT IB2013, 002233.
\bibitem[Choi et al.(1999)]{choi1999}Choi, H. S., Kim, S. L., Kang, Y. H., Park, B. G., 1999. Search for Variable Stars in the Open Cluster NGC7654, A\&A 348, 789.
\bibitem[Chonis and Gaskell(2008)]{chonis2008}Chonis, T. S., Gaskell, C. M., 2008. Setting $UBVRI$ Photometric Zero-Points Using Sloan Digital Sky Survey $ugriz$ Magnitudes, AJ 135, 264.
\bibitem[Cox(2000)]{cox2000}Cox, A. N., 2000. Allen's Astrophysical Quantities (4th ed.; Los Alamos:Springer).
\bibitem[de Wit et al.(2006)]{de wit2006}de Wit, W. J., Bouvier, J., Palla., F., et al., 2006. Exploring the Lower Mass Function in the Young Open Cluster IC4665, A\&A 448, 189.
\bibitem[Eggen(1971)]{eggen1971}Eggen O. J., 1971. Luminosities, Temperatures, and Kinematics of K-type Dwarfs, ApJS 22, 389.
\bibitem[Fukugita et al.(1996)]{fukugita1996}Fukugita, M., Ichikawa T., Gunn, J. E., et al., 1996. The Sloan Digital Sky Survey Photometric System, AJ 111, 1748.
\bibitem[Haug(1970)]{haug1970}Haug, U., 1970. $UBV$ Observations of Luminous Stars in Three Milky Way Fields, A\&AS 1, 35.
\bibitem[Henden et al.(2014)]{henden2014}Henden, A., Turner, R., Kloppenborg, B., et al., 2014. The AAVSO DSLR Observing Manual, Report for AAVSO. ISBN 978-1-939538-07-9, pp. 35.
\bibitem[Hogg and Kron(1955)]{hogg1955}Hogg, A. R., Kron, G. E., 1955. The Galactic Cluster IC4665, AJ 60, 365.
\bibitem[Hoot(2007)]{hoot2007}Hoot, J. E., 2007. Photometry with DSLR Cameras, SASS 26, 67.
\bibitem[Jeffries(2009)]{jeffries2009}Jeffries, R. D., Jackson, R. J., James, D. J., Cargile, P. A., 2009. Low-mass Members of the Young Cluster IC 4665 and Pre-main-sequence Lithium Depletion, MNRAS 400, 317.
\bibitem[Johnson(1954)]{johnson1954}Johnson, H. L., 1954. The Standard Region Near IC4665, ApJ 119, 181.
\bibitem[Karaali et al.(2005)]{karaali2005}Karaali, S., Bilir, S., Tun\c{c}el, S., 2005. New Colour Transformations for the Sloan Photometry, and Revised Metallicity Calibration and Equations for Photometric Parallax Estimation, PASA 22, 24.
\bibitem[Karaali and Yaz G\"{o}k\c{c}e(2013)]{karaali2013}Karaali, S., Yaz G\"{o}k\c{c}e, E., 2013. Metallicity-Dependent Transformations for Red Giants with Synthetic Colours of $UBV$ and $ugr$, PASA 30, 40.
\bibitem[Kloppenborg et al.(2012)]{kloppenborg2012}Kloppenborg, B. K., Pieri, R., Eggenstein H. -B., et al., 2012. A Demonstration of Accurate Wide-field V-band Photometry Using a Consumer-grade DSLR Camera, JAAVSO 40, 815.
\bibitem[Landolt(1983a)]{landolt1983a}Landolt, A. U., 1983a. $UBVRI$ Photometric Standard Stars Around the Celestial Equator, AJ 88, 439.
\bibitem[Landolt(1983b)]{landolt1983b}Landolt, A. U., 1983b. $UBVRI$ Photometry of Stars Useful for Checking Equipment Orientation Stability, AJ 88, 853.
\bibitem[Landolt(1992)]{landolt1992}Landolt, A. U., 1992. Broadband $UBVRI$ Photometry of the Baldwin-Stone Southern Hemisphere Spectrophotometric Standards, AJ 104, 372.
\bibitem[Lodieu et al.(2011)]{lodieu2011}Lodieu, N., de Wit, W. -J., Carraro, G., et al., 2011. The Mass Function of IC4665 Revisited by the UKIDSS Galactic Clusters Survey, A\&A 532, 103.
\bibitem[Loughney(2010)]{loughney2010}Loughney, D., 2010. Variable Star Photometry with a DSLR Camera, JBAA 120, 157.
\bibitem[Maciejewski and Niedzielski(2007)]{maciejewski2007}Maciejewski, G., Niedzielski, A., 2007. CCD $BV$ Survey of 42 Open Clusters, A\&A 467, 1065.
\bibitem[Manzi et al.(2008)]{manzi2008}Manzi, S., Randich, S., de Wit, W. J., Palla, F., 2008. Detection of the Lithium Depletion Boundary in the Young Open Cluster IC4665, A\&A 479, 141.
\bibitem[McCarthy and O'Sullivan(1969)]{mccarthy1969}McCarthy, M. F., O'Sullivan, S., 1969. A Photometric Study of the Open Cluster IC4665, RA 7, 483.
\bibitem[Menzies et al.(1991)]{menzies1991}Menzies, J. W., Marang F., Laing, J. D., Coulson, I. M., Engelbrecht, C. A., 1991. $UBV(RI)_C$ Photomety of Equatorial Standard Stars. A Direct Comparison Between the Northern and Southern Systems, MNRAS 248, 642.
\bibitem[Menzies and Marang(1996)]{menzies1996}Menzies, J. W., Marang, F., 1996. $UBV(RI)_C$ Observations of Johnson's Standard Sequence in IC4665, MNRAS 282, 313.
\bibitem[Muzzio(1973)]{muzzio1973}Muzzio, J. C., 1973. $UBV$ Photometry of Large Proper Motion Stars, PASP 85, 358.
\bibitem[Neckel(1974)]{neckel1974}Neckel, H., 1974. Photoelectric Catalogue of 1030 BD M-type Stars Located a Long the Galactic Equator, A\&AS 18, 167.
\bibitem[O'Donnell(1994)]{odonnell1994}O'Donnell, J. E., 1994. $R_V$-Dependent Optical and Near-Ultraviolet Extinction, ApJ 422, 158.
\bibitem[Pandey et al.(2001)]{pandey2001}Pandey, A. K., Nilakshi, Ogura, K., Sagar, R., Tarusawa, K., 2001. NGC7654: An Interesting Cluster to Study Star Formation History, A\&A 374, 504.
\bibitem[Pesch(1960)]{pesch1960}Pesch, P., 1960. The Galactic Cluster NGC 7654 (M52), ApJ 132, 689.
\bibitem[Prosser(1993)]{prosser1993}Prosser, C. F., 1993. The Open Cluster IC4665, AJ 105, 1441.
\bibitem[Rodgers et al.(2006)]{rodgers2006}Rodgers, C. T., Canterna, R., Smith, J. A., et al., 2006. Improved $u^\prime g^\prime r^\prime i^\prime z^\prime$ to $UBVR_CI_C$ Transformation Equations for Main-Sequence Stars, AJ 132, 989.
\bibitem[Sanders and van Altena(1972)]{sanders1972}Sanders, W. L., van Altena, W. F., 1972. Membership of the Open Cluster IC4665, A\&A 17, 193.
\bibitem[Schlegel et al.(1998)]{schlegel1998}Schlegel, D. J., Finkbeiner, D. P., Davis, M., 1998. Maps of Dust Infrared Emission for Use in Estimation of Reddening and Cosmic Microwave Background Radiation Foregrounds, ApJ 500, 525.
\bibitem[Smith et al.(2002)]{smith2002}Smith, J. A., Tucker, D. L., Kent, S., et al., 2002. The $u^\prime g^\prime r^\prime i^\prime z^\prime$ Standard-Star System, AJ 123, 2121.
\bibitem[Smith et al.(2011)]{smith2011}Smith, R., Jeffries, R. D., Oliveira, J. M., 2011. Debris Discs in the 27 Myr Old Open Cluster IC4665, MNRAS 411, 2186.
\bibitem[Stencel(2013)]{stencel2013}Stencel, R. E., 2013. Results of the Recent $\varepsilon$ Aurigae Eclipse Campaign, CEAB 37, 85.
\bibitem[Stetson(2000)]{stetson2000}Stetson, P. B., 2000. Homogeneous Photometry for Star Clusters and Resolved Galaxies. II. Photometric Standard Stars, PASP 112, 925.
\bibitem[Tiffany(2008)]{tiffany2008}Tiffany, M. B., 2008. The Search for Transiting Extrasolar Planets in the Open Cluster M52, SDSU, pp. 10.
\bibitem[Tweet et al.(2009)]{tweet2009}Tweet, D. J., Lee, J. J., Speigle, J. M., Tamburrino, D., 2009. 2PFC CMOS Images Sensors: Better Image Quality at Lower Cost, SPIE 7250, 7205007.
\bibitem[Viskum et al.(1997)]{viskum1997}Viskum, M., Hernandez, M. M., Belmonte, J. A., Frandsen, S., 1997. A Search for Delta Scuti Stars in Northern Open Clusters I. CCD Photometry of NGC7245, NGC7062, NGC7226 and NGC7654, A\&A 328, 158.
\bibitem[V\'{i}tek and Bla\u{z}ek(2012)]{vitek2012}V\'{i}tek, S., Bla\u{z}ek, M., 2012. Notes on DSLR Photometry, ASI Conf. Series 7, 231.
\bibitem[Wramdemark(1976)]{wramdemark1976}Wramdemark, S., 1976. Distant Early-type Stars in Some Northern Milky Way Fields, A\&AS 26, 31.
\bibitem[Yaz et al.(2010)]{yaz2010}Yaz, E., Bilir, S., Karaali, S., et al., 2010. Transformations Between the 2MASS, SDSS and $BVI$ Photometric Systems for Late-type Giants, AN 331, 807.

\end{thebibliography}
\end{document}